# Direct and full-scale experimental verifications towards ground-satellite quantum key distribution

Jian-Yu Wang[1,2,†], Bin Yang[1,†], Sheng-Kai Liao[1,2], Liang Zhang[2], Qi Shen[1], Xiao-Fang Hu[1], Jin-Cai Wu[2], Shi-Ji Yang[2], Hao Jiang[2], Yan-Lin Tang[1], Bo Zhong[3], Hao Liang[1], Wei-Yue Liu[3], Yi-Hua Hu[2], Yong-Mei Huang[4], Bo Qi[4], Ji-Gang Ren[1], Ge-Sheng Pan[1], Juan Yin[1], Jian-Jun Jia[2], Yu-Ao Chen[1], Kai Chen[1], Chen-Zhi Peng[1], and Jian-Wei Pan[1]

[1] Shanghai Branch, National Laboratory for Physical Sciences at Microscale and Department of Modern Physics, University of Science and Technology of China, Shanghai, 201315, P.R. China
[2] Shanghai Institute of Technical Physics, Chinese Academy of Sciences, Shanghai, 200083, P.R. China
[3] College of Information Science and Engineering, Ningbo University, Ningbo, 315211, P.R. China
[4] The Institute of Optics and Electronics, Chinese Academy of Sciences, Chengdu 610209, P.R. China
[†] These authors contributed equally to this work

**Abstract: Quantum key distribution (QKD)[1,2] provides the only intrinsically unconditional secure method for communication based on principle of quantum mechanics. Compared with fiber-based demonstrations[3-5], free-space links could provide the most appealing solution for much larger distance. Despite of significant efforts[6-13], so far all realizations rely on stationary sites. Justifications are therefore extremely crucial for applications via a typical Low Earth Orbit Satellite (LEOS). To achieve direct and full-scale verifications, we demonstrate here three independent experiments with a decoy-state QKD system overcoming all the demanding conditions. The system is operated in a moving platform through a turntable, a floating platform through a hot-air balloon, and a huge loss channel, respectively, for substantiating performances under rapid motion, attitude change, vibration, random movement of satellites and in high-loss regime. The experiments cover expanded ranges for all the leading parameters of LEOS. Our results pave the way towards ground-satellite QKD and global quantum communication network.**

Quantum key distribution (QKD)[1, 2] appears as the first important application among the evolving field of quantum information technology. Currently the maximum distance for practical QKD in fiber has achieved around an order of ~ 100 km[3-5], which almost reaches its limit with state-of-the-art technology due to huge photon loss, noise of available single photon detector, and decoherence effect in the optical fiber[14]. Due to low absorption and negligible nonbirefringent character at atmosphere, optical free space therefore serves as the most promising channel for large-scale quantum communication by use of satellites and ground stations[15-17]. Consequently secret keys can be established between any two sites globally.

There have been significant theoretical progresses for feasibility of ground-satellite quantum communications, along with some preliminary experimental tests. It is shown by Hughes *et al.* that daylight and nighttime operation is accomplished for practical free-space QKD over a 10 km path[7], while the first experimental quantum entanglement distribution over 13 km is reported[8]. Recently the long distance of 144 km experiment have been successfully performed by Zeilinger's group[10-12], which reveals a fact that long-distance atmospheric turbulence has no much effect on quantum communications. Moreover entanglement QKD is illustrated for free-space optical link over 1.5 km[9]. Besides, quantum teleportation has sequentially been realized in real world atmospheric conditions over 16 km[13] and 100 km[18, 19]. These experiments have accumulated indispensable technical advances, and have formed a solid basis for free-space based quantum communication. Some of the theoretical aspects are also derived[15, 16, 20, 21] for feasibility of ground-satellite QKD.

In a real-life situation, however, quantum communication based on satellite needs to handle several critical issues: the first issue is that the satellite has a rapid relative angular motion with regard to ground stations; the second issue is that the satellite may have unwanted random motion, while the third issue is to overcome atmospheric turbulence and to generate secret keys under condition of high-loss regime. For a typical 400 km~800 km LEOS, the motion parameters are generally as follows: maximum angular velocity at 20 mrad/s, maximum angular acceleration at 0.23 mrad/s$^2$. When the orbit goes higher, the angular velocity and angular acceleration will be smaller. If equipped a receiving telescope with diameter of around 1 m, direct estimation shows that the loss is about 30 dB~50 dB via a channel linking ground station and LEOS. To put into practice of satellite-based QKD, direct and full-scale verifications are therefore of paramount importance to overcome all of these real-life issues for extracting successfully secret keys through rapid moving and vibration sites, and against high-loss circumstance. This would enable to build steady quantum channel link, maintain maximum possible channel efficiency, reduce quantum bit error rate (QBER), and achieve high signal-to-noise ratio substantially. However, we remark that it is impossible to simultaneously combine the three issues in a single experiment at the moment except in a real satellite. On the one hand, it is a challenging task to find a platform of aero craft that could have simultaneously big angular velocity and angular acceleration. On the other hand, combining further with high loss, random motion and attitude change is impossible in any available platforms. Take an air craft for example, it provides neither suitable angular acceleration together with big angular velocity, nor high loss environment, random motion and attitude change with a degree as high as a satellites, at the same time.

To address every aspect of the above mentioned crucial issues, here we report direct and comprehensive verifications for establishing successful quantum cryptography communication via satellites, through three independent experiments at night. Firstly quantum communication experiments over a turntable and a hot-air balloon, respectively, to simulate the platform of rapidly moving orbit, as well as vibration, random motion and attitude change for satellite. Afterwards we illustrate generation of secret keys for a 96 km free-space channel with about 50 dB loss, which is severe than the case of 30 dB~50 dB loss for link between ground stations and LEOS[15, 16]. A high-speed QKD[22-28] system based on decoy scheme is developed and testified under each scenario of these real-life situations. Furthermore, a system of acquisition, tracking and pointing (ATP) is designed to accomplish coarse and fine tracking, for establishing steady and accurate connection of optical links. High-speed optical source and controlling electronics are developed to operate at 100 MHz. Tailored transmitters and receivers are engineered and integrated to be slight and portable terminals. Low background counts and high temporal precision are maintained to achieve high signal-to-noise ratio. Besides to compensate the polarization-basis deflection, a three-dimensional platform is set up for transmitter terminal, which is another critical ingredient for satellite communications based on polarization encoding. The overall system has successfully maintaining quantum key distribution process based on turntable through public free-space channel of 40 km, via a floating hot-air balloon of 20 km, and over a 96 km link with about 50 dB loss, respectively. The secure distances achieved are significantly longer than the effective atmosphere thickness (equivalent to 8~10 km of ground atmosphere). Our verification environment has not only incorporated all the possible motion modes, but also consisted of more extreme situations like vibration, random movement, attitude change for satellites through hot-air balloon operating. Therefore our implementations, for the first time, provide comprehensive and direct verifications for secure key exchanges through fast-moving platforms like satellite or aircraft, and make closer for a global scale quantum communication.

A schematic layout of the moving and floating platform experimental setup is shown in Fig. 1. Our experiment employs the decoy-state protocol proposed by Hwang[22], developed by Wang[23] and by Lo *et al.*[24, 25] independently and systematically to achieve unconditional security for practical applications. We manage to design and integrate the optical components, electronics, and telescope for the transmitter and the receiver to be slight and portable terminals. In the transmitter terminal, the optical source contains four laser diodes which emit 1 ns optical pulses centered at 850 nm with a full width at half maximum (FWHM) of 0.5 nm. Polarization encoded module is connected to optical source through single-mode fibers which consist of two polarizing beam splitters (PBS), one beam splitter (BS) and one half-wave plate (HWP). The optical pulses emitted from the polarization encoded module are four polarization states of $|H\rangle$, $|V\rangle$, $|+\rangle$, $|-\rangle$, where $|H\rangle$, $|V\rangle$ represent horizontal polarization and vertical polarization, $|+\rangle = (1/\sqrt{2})(|H\rangle + |V\rangle)$ and $|-\rangle = (1/\sqrt{2})(|H\rangle - |V\rangle)$, as the four states for the standard BB84 protocol. The type and the amplitude of random pulsed signals emitted from the four diodes are all controlled according to high-speed random numbers generated beforehand by a random physical noise (RND). In order to make the amplitude of optical pulses identical and to monitor deviation of optical power in real-time, another export of the BS in the polarization encoded module is connected to power meter as a monitor window (MON). After the polarization encoded module is an attenuator (ATT) that is used to attenuate the average number of per pulse to the experimental level. Then a BS, a fast steering mirror (FSM) and a Complementary Metal Oxide Semiconductor (CMOS) constitute a fine tracking system. The BS is used to collect the 671 nm beacon light from receiver to fine tracking CMOS, while FSM is used to dynamically adjust optical path according to a correction program through the image information obtained by the fine tracking CMOS. Between the BS and fine tracking CMOS, there is a dichroic mirror (DM) to transmit 671 nm beacon light into CMOS and reflect 532 nm pulsed synchronization light emitted from 532 nm laser. The 532 nm pulsed light with optical power of 100 mw and divergence angle of 1 mrad acts as also the beacon light of receiver system. We have engineered the above mentioned optical components into a luggable knapsack. The primary mirror of the transmitter terminal is a reflecting Cassegrain telescope with aperture of 200 mm, focal length of 1250 mm, and magnification of 10, which has small volume and fits well our experimental requirements. The coarse tracking CMOS are set at the top of telescopic tube. The tracking system contains coarse tracking and fine tracking, whose principles and performances are described in the Methods section.

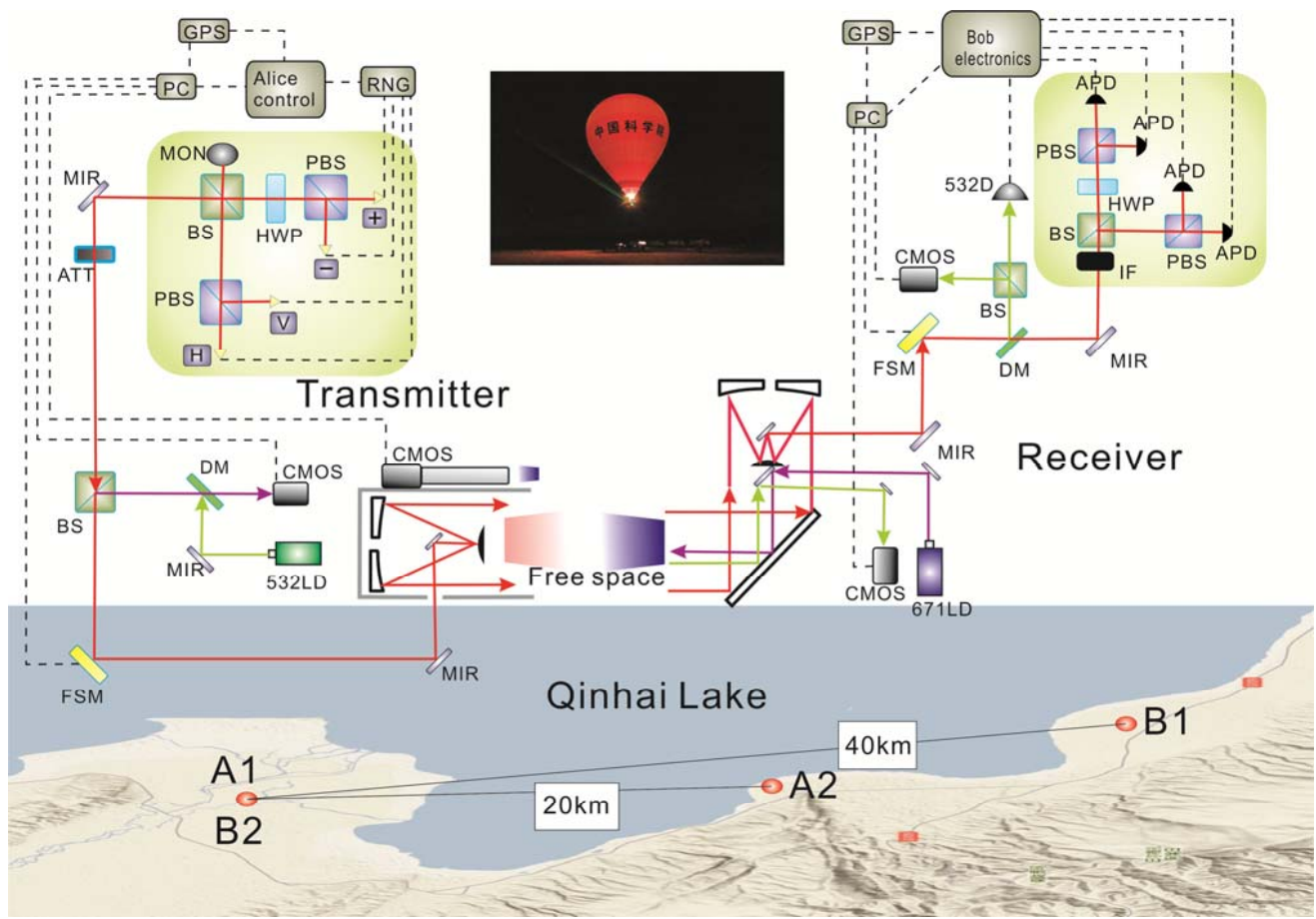

**Figure 1: Schematic diagram of the experimental setup.** The signal states and decoy states emitted from polarization encoded module go through reflecting Cassegrain telescope before transmitting to the receiver side. Once received by another telescope, they are directed to the detection module for polarization analysis. The 532 nm and 671 nm beacon lights pass through the same channel for tracking,
5

while the 532 nm light also acts as synchronization signal. A1 (A2) and B1 (B2) represent the transmitter's and receiver's sites for the moving (floating) platform experiments, respectively. Red lines represent the light of 850 nm, green lines and violet lines represent the tracking beacon light of receiver and transmitter, respectively, while dotted lines represent the electric cable. PBS: polarizing beam splitters; BS: beam splitter; HWP: half-wave plate; MON: monitor window; MIR: mirror; ATT: Attenuator; DM: dichroic mirror; 532LD: 532 nm laser; FSM: fast steering mirror; 671LD: 671 nm laser; 532D: 532 nm detector; IF: interference filter; APD: avalanche photo diode. The inset shows rising and erupting hot-air balloon in the experiment of floating platform.

As for the receiver terminal, the primary mirror is a reflecting Cassegrain telescope with aperture of 300 mm, focal length of 1500 mm, and magnification of 12. The optical components of receiver are set to a small optical table. The tracking system also contains coarse tracking and fine tracking, whose performance parameters are listed in the Method section. The light collected contains 532 nm beacon light, 850 nm signal and decoy light. We remark that part of the 532 nm beacon light is directly reflected to the coarse tracking CMOS, while the beacon light of 671nm with optical power of 300 mw goes to the telescope before flying to the transmitter (see Fig. 1). The light out from telescope firstly passes through the fine tracking system consisted of a FSM, a DM, and a BS. The FSM is used for fine tracking's feedback, while the DM (transmitting 850 nm and reflecting 532 nm) is employed to separate the light of 532 nm from 850 nm. The light 532 nm is then divided into two parts by the BS for fine tracking and synchronization, respectively. Transmitting light after the DM goes through a closed receiver module for polarization analysis. Different from the encoding module using single mode fiber, here the connector deploys multimode fiber with core diameter of 105 $\mu$m. Before the entrance of receiver module there is an interference filter (IF) (center wavelength at 850 nm, 5 nm FWHM) which could greatly reduce the effects of stray light. Collected light from multimode fiber are led to the avalanche photodiodes (APD). The receiver electronics acquires channel and timing data of APD and the timing data of the synchronization light, before transferring to a personal computer (PC) for further processing. Concerned with synchronization, after using GPS for an initially coarse synchronization we have employed an approach of external synchronization with the help of 532 nm pulsed light, and have reached an accuracy of 1 ns between the distant transmitter and receiver.

During the experiment, we have first adjusted the intensities to be identical for the signal states from all the laser diodes. The decoy states intensities are adapted to be the same as well. The average photon number of signal states and decoy states per pulse is reduced to be around 0.8 and 0.27 (0.6 and 0.2 in high-loss experiment) by suitable attenuating. In order to erase the temporal information, every length of optical fiber, length of cable, and optical path are kept to be uniform. In the experiment, the total repetition rate of signal-state, decoy-state and vacuum-state is 100 MHz, with a proportion of 2:1:1 for the three types of states. The power of 850 nm optical source is monitored in real-time, and the experiment would be terminated if the power fluctuates over 5%.

In the first experiment, we choose Administration Bureau of Qinghai Lake National Nature Reserve（N37°2'16" E99°44'33"） for transmitter (Alice) and Heimahe Nature Reserve Station for receiver (Bob)（N36°42'1" E99°52'16"） in the experiment of satellite orbiting simulation, with a straight-line distance between transmitter and receiver of about 40 km. The experiment was performed at the end of August 2010. We place the transmitter terminal on a turntable of type Aerotech AGR-150, with the maximum torque of 4 Nm. The turntable exhibits a complex nonlinear mode for relative motion between transmitter and receiver, which amounts to a motion with a maximum angular velocity of 21 mrad/s and a maximum angular acceleration of 8.7 mrad/s$^2$. The motion realm has covered completely the possible motion range parameters for a 400 km~800 km LEOS. In the receiver site, the APDs (total background counts of 800/s) could collect a rate of about 5000/s, which refers to the total loss of about 40 dB. We contribute 19 dB due to geometric attenuation, 6 dB due to atmospheric loss, 13 dB due to optical system of receiver (3 dB from optical elements, 3 dB from detector efficiency, and 7 dB from coupling efficiency of receiver module), and an extra 1 dB~5 dB due to the efficiency decreased by moving channel link maintained by ATP.

To further demonstrate the applicability of our experimental set-up and verify the feasibility of satellite QKD in the vibration and floating platform, we have implemented the QKD system in floating hot-air balloon, which is for the first time a QKD experiment with random motion, attitude change and vibration simulations. On September of 2010, we set up the transmitter terminal at a hot-air balloon, whose location is a grass of Qinghai lakeside (N36°49'43" E99°44'18"). The receiver is placed at bird island hotel (N37°2'16" E99°44'31") of Qinghai Lake, with a straight-line distance of about 20 km from the hot-air balloon. The essential part of the experiment is to build a steady optical link between the transmitter and the receiver. When the hot-air balloon rises, it will have random motion such as rotating, jolting, and shaking, which is more rigor than the possible random motion, attitude change and vibration of satellite. After monitoring and recording, an estimate gives an average angular velocity of 10.5 mrad/s, and an average angular acceleration at 1.7 mrad/ s$^2$ for random motion of a hot-air balloon. Under the case the fine tracking accuracy is below 5 $\mu$ rad, which is much higher than requirement of 10 $\mu$ rad for a typical LEO. We manage to choose the rising hot-air balloon in moorage state such that coarse tracking works within the field of view. This is attained by adjusting and fastening the ropes from four directions when the rising hot-air balloon has drastic motion. Therefore the ATP's ability for recapturing is very important. In our experiment when the balloon makes dramatic motion, the system will be out of view. However, the system can still recapture the target and generate keys rapidly, typically within 3~5 seconds. A successful link will be fleetly reconstructed in the case, which shows superior ability for the ATP. This situation indicates that our ATP could satisfy requirement for satellite-based QKD. To ensure the whole system working well, we have firstly carried out a ground test of free-space decoy-state QKD before rising of balloon. Another factor of reducing effective experimental time is the fact that the balloon has to erupt flame every few seconds. This will cause more stray counts, so that the valid data of time will be less. Currently our experiment is favorable under good weather conditions such as no rain, good atmosphere and wind speed less than 3 m/s.

To verify whether one can really attain high signal-to-noise rate and overcome the obstacle under high-loss environment of satellite QKD, we implement also an experiment of long distance free-space QKD with about 50 dB loss between stationary sites. The experiment was performed around October of 2010. The transmitter (Alice) is located in China, Qinghai, Hainan autonomous prefecture, Qinghai lake observatory (North Latitude $36°33'17"$, East Longitude $100°28'30"$, elevation 3585m） and receiver (Bob) is chose in China, Qinghai, Haibei autonomous prefecture,

Quanji village (North Latitude $37°16'44''$, East Longitude $99°53'4''$, elevation 3255m). Straight-line distance between transmitter and receiver is about 96 km. The main difference with the moving and floating platform experiments is the receiver, which is a reflecting Cassegrain telescope with effective aperture of 600 mm, focal length of 6900 mm, and magnification of 5.6. We choose the avalanche photodiodes (APDs) of PerkinElmer SPCM-AQRH-16 with dark counts of 25/s and detect efficiency of above 45% to reduce the dark counts. The other devices are the same with the previous. The average photon number per pulse of signal state and decoy state is, 0.6 and 0.2, respectively. We receive about 500 cps counts with total background counts of 120 cps (dark counts of four APD is about 100 cps while stray counts only 20 cps), so our experimental total loss is as high as 50 dB. Next we analyze the attenuation of the link: divergence angle of transmitter telescope is 80 $\mu$rad (it is amplified by atmosphere while its measure value of 40 $\mu$rad in the lab), so geometric attenuation of 96 km free space is about 22 dB; atmospheric attenuation of 96 km free space is about 8 dB; the transmittance of receiving telescope is 50%, efficiency of receiver optical component is 80%, detector efficiency is 45%, receiver coupler efficiency is 10%, these items total attenuation is about 17.5 dB, effect on ATP and atmospheric turbulence is about 2-5 dB, so total loss is over 50 dB which matched the counts we received.

From the experimental parameters we can obtain each counting rate and the error rate for all the three types of states. The final secure key rate that can be distilled through the systematic results[22-25, 29, 30]:

$$R \geq q\{-Q_\mu f(E_\mu) H_2(E_\mu) + Q_1[1 - H_2(e_1)]\},$$

where the subscript $\mu$ is the average photon number per signal pulse. For convenience, we denote $\nu$ the average photon number per decoy pulse. $q$ is an efficiency factor for the protocol; $E_\mu$ and $Q_\mu$ are the QBER and the measured gain for signal states, respectively. $e_1$ and $Q_1$ are the error rate and the unknown gain of the true single photon state in signal states. To achieve maximum possible key generation rate, the decoy-state method can estimate the upper bound of $e_1$ denoting as $e_1^U$, and the lower bound of $Q_1$ denoting as $Q_1^L$. The $H_2(x)$ is the binary entropy function: $H_2(x) = -x\log_2(x) - (1-x)\log_2(1-x)$, while the factor $f(x)$ is for considering an efficiency of the bi-directional error correction.

We follow here the methods developed independently and systematically by Wang[23] and by Lo et al.[24] for generating unconditional secure keys. After experimentally measuring all the relevant parameters as listed in Table 1, and considering finite key length as well as statistical fluctuations, we can input the following bounds for calculating final security key generation rate[25]

$$Q_1 \geq Q_1^L = \frac{\mu^2 e^{-\mu}}{\mu v - v^2}(Q_v^L e^v - Q_\mu e^\mu \frac{v^2}{\mu^2} - Y_0^U \frac{\mu^2 - v^2}{\mu^2}),$$

$$e_1 \leq e_1^U = \frac{E_\mu Q_\mu - Y_0^L e^{-\mu}/2}{Q_1^L},$$

in which

$$Q_v^L = Q_v(1 - \frac{10}{\sqrt{N_v Q_v}}),$$

$$Y_0^L = Y_0(1 - \frac{10}{\sqrt{N_0 Y_0}}),$$

$$Y_0^U = Y_0(1 + \frac{10}{\sqrt{N_0 Y_0}}).$$

Here $N_v$ and $N_0$ are numbers of pulses used as decoy states and vacuum state, respectively, while $Q_v$ is the measured gain for the decoy states. The measured counting rate for vacuum decoy states is denoted by $Y_0$.

**Table 1 Experimental parameters and results**

| P | Moving platform over a turntable | Floating platform (a hot-air balloon) | High loss channel (96 km, 50 dB loss) |
|---|---|---|---|
| $N$ | $2.7*10^{10}$ | $3.15*10^{10}$ | $1.59*10^{11}$ |
| $Q_\mu$ | $5.34*10^{-5}$ | $5.90*10^{-5}$ | $4.41*10^{-6}$ |
| $Q_v$ | $2.17*10^{-5}$ | $2.42*10^{-5}$ | $2.36*10^{-6}$ |
| $Y_0$ | 3.00*10<sup>-6</sup> | 2.69*10<sup>-6</sup> | 4.30*10<sup>-7</sup> |
| $E_\mu$ | 2.73% | 2.35% | 4.04% |
| $R_\mu$ | 6.38*10<sup>-6</sup> | 1.08*10<sup>-5</sup> | 1.92*10<sup>-6</sup> |
| $K_\mu/T$ | 159.41 | 268.87 | 48.03 |

**$N$ is the number of total pulses, $Q_{\mu(\nu)}$ is the counting rate that comes from signal states (decoy states), $Y_0$ is the counting rate from vacuum states, $E_\mu$ denotes QBER of the signal states. $K_\mu$ is the total final secret keys distilled from raw keys.**

The original experimental data and derived results are listed in Table 1. Based on moving turntable, the decoy-state QKD is successfully accomplished over 40 km free-space. The final secure keys of 43038 bits are generated in 270 s, which amounts to a key rate of 159.4 bit/s. Please note that in the floating experiment, random movement of balloon influence from wind and airflow would impose significant challenge for normal QKD system. Fortunately, with the help of ATP system, integrated designing of optical system and controlling electronics, we have maintained successful operating of QKD under such conditions. Final secure keys are distributed with a rate of 268 bit/s in 315s and a quite low QBER of 2.35%. This confirms that our system of QKD is robust and resistant to relative adverse circumstances. In the experiment of 96 km QKD, we distill secret keys with the key rate of 48 bit/s and quantum bit error rate of 4.04% from which 1% due to polarization deflection of devices, 2.2% due to dark counts and the others due to temporal precision. Our results show a common feature of low QBER, which due to our endeavor on the polarization maintaining optics, high temporal precision and low background counts. The cumulative effects of timing jitter of the Alice electronics, GPS noise, timing jitter of the detector, timing jitter of the TDC and other temporal noise lead to a total temporal precision of signal events with a FWHM of 1.1 ns. For raw key generation, we take time window of 2 ns by the software, leading to further reduction of background counts (contain dark counts).

In conclusion, we have achieved significant experimental results towards the realization of QKD in free space via quantum channel from satellites to ground stations. Three independent experiments are performed successfully, including over a rapidly moving platform on distance of 40 km, a floating platform with distance of 20 km, and over 96 km air with a huge loss. Our presentations provide directly complete and comprehensive verifications for the real scenarios of quantum communication via satellite or aircraft, which accompanies with moving, vibrating terminals with possibly random motion as well as attitude change. We also show that atmospheric turbulence and huge loss circumstances could be overcome in satellite QKD, as confirmed by demonstration undergoing much more severe conditions compared with real and typical LEOS. Various techniques, particularly precise tracking through the self-made ATP system, and the integrated communication terminals etc., are developed to maintain smooth distribution of secret keys. Emitted from moving objects, high quality quantum signals are successfully extracted after flying a significant distance through air. In both moving platform and floating platform cases, key rates of greater than 150 bit/s are reached, with low QBER less than 2.8%. When the distance of QKD is expanded to 96 km, one can finally maintain the key rates of 48 bit/s associated with low QBER about 4%. The achieved distance between communication parties exceeds several times of the effective atmosphere thickness. Our results represent the first milestone on the road to direct and full-scale experimental verifications towards satellite-based quantum communication, and particularly have solved the key issues for building steady optical links with rapidly moving, attitude changing, vibrated terminals, as well as high loss environment. The technical advances are readily applicable to satellite based distribution of quantum keys, and suggest methods of global scale quantum communication network in real-life scenarios.

*Note added:* After submission of our manuscript, we become aware of a similar experiment claimed in Ref. [31] using an airplane for QKD transmission with the BB84 protocol.

**METHODS**:

**ATP**:

The acquisition, tracking and pointing (ATP) system is necessary for building the link of free-space quantum key distribution. The ATP system is composed of two control loops: coarse control loop and fine control loop. The coarse control loop is constructed with coarse pointing mechanism (CPM), coarse controller and coarse camera. The fine control loop is constructed with fine pointing mechanism (FPM), fine controller and fine camera. The coarse camera has a wide field of view, while the fine camera has a small field of view. First, based on the open pointing of CPM, the coarse camera captures beacon light from the target. Then CPM corrects its pointing direction based on the position of the light spot in the coarse camera and guides the beacon light into the fine camera. The FPM corrects its pointing direction base on the light spot position from the fine camera. The coarse control loop has a tracking accuracy of several hundred microradians while the fine control loop decreases the tracking error to several microradians. The performance of the ATP systems in our experiment is listed in Table 2.

Table 2 Performance of the ATP system

| Components | | Transmitter terminal | Receiver terminal |
|---|---|---|---|
| Telescope diameter | | 200 mm | 300 mm |
| Coarse Pointing Mechanism | Type | Two-axis gimbal mount | Two-axis gimbal mirror |
| | Tracking Range | Azimuth:±45°; Elevation:±70° | Both : ±5° |
| Coarse Camera | Field of View | 2° | 1° |
| | Size | 1000×1000 pixels | 640×480 pixels |
| Fine Pointing Mechanism | Type | Fast steering mirror | Fast steering mirror |
| | Tracking Range | ±0.7 mrad | ±0.7 mrad |
| Fine Camera | Field of View | 512 $\mu$rad | 512 $\mu$rad |
| | Size & Frame per Second | 128×128 & 2300Hz | 128×128 & 2300Hz |
| Tracking Errors | Coarse Tracking Error | ±200 $\mu$rad | |
| | Fine Tracking Error | ±5 $\mu$rad | |

In the experiment of satellite orbiting simulation, the turntable would bring additional disturbance especially when it turns to another direction during the movement. So ATP would have some obvious return error. The coarse tracking error is therefore below ±200 $\mu$rad by getting rid of these return errors, while fine tracking error is about ±5 $\mu$rad (RMS), as illustrated in Fig. 2. While in the experiment of simulating the vibration of satellite, hot-air balloon sometimes would have severely movement and make the beacon out of the camera field, nevertheless our ATP system could recapture rapidly. When stabilized, the tracking accuracy is about ± 5 $\mu$ rad (RMS), as shown in Fig. 3. The control bandwidth of fine tracking is above 200 Hz, while that the achieved disturbance rejection bandwidth is above 100 Hz[18]. It should be remarked that in the Fig. 2, time axis (x axis) seems not to be continuous, which is due to a processing of combination of several sets of data. Thus the period of time axis is not exactly the turntable period of 15 s. This phenomenon only comes from the limitation of the date storage of CMOS, and does not influence experimental results.

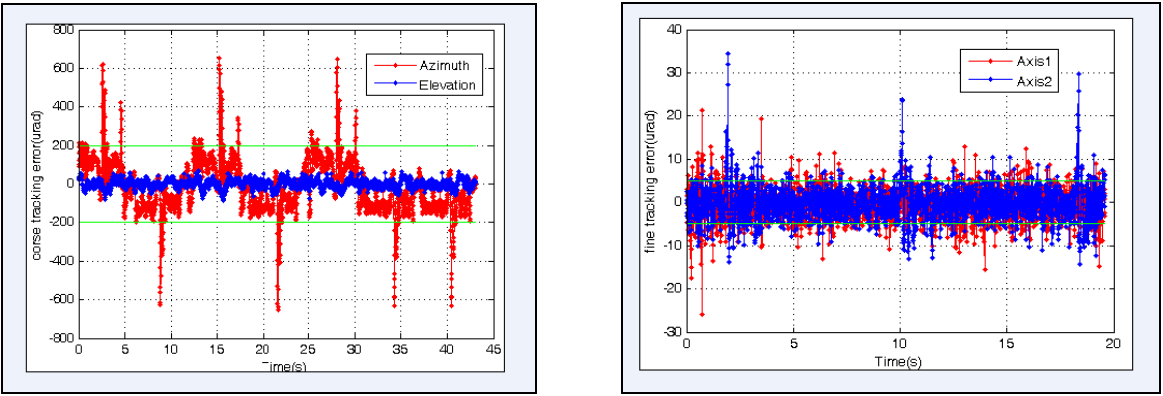

(a)                                             (b)

Figure 2: Tracking error of the ATP system in the satellite orbiting simulation by a turntable. **a,** The coarse tracking error of two axies. **b,** The fine tracking error at the same time. The coarse error is between ±200 $\mu$ rad after getting rid of return error from the moving table, while the fine tracking error is about ±5 $\mu$ rad (RMS).

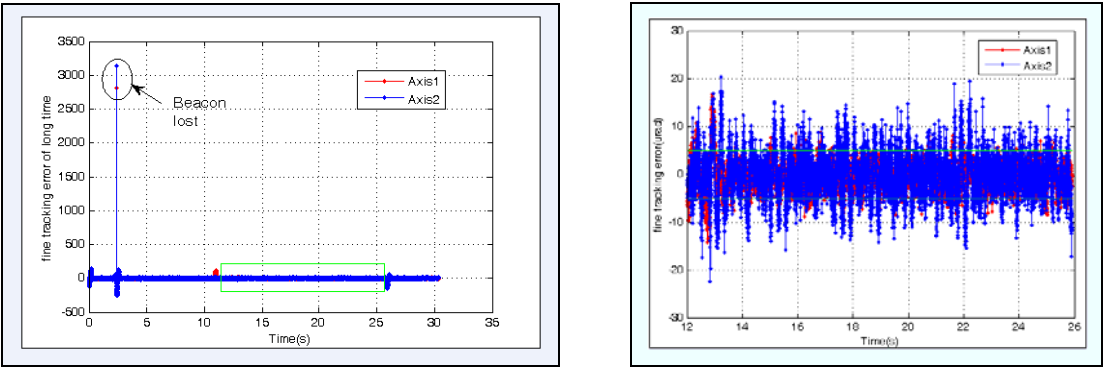

(a)                                             (b)

Figure 3: Tracking error of the ATP system in the vibration of satellite simulation by a hot-air balloon. Figure (a) is a long time fine tracking error with beacon light out of field of view sometimes. Figure (b) is the fine tacking error in the stabilized time area. When the beacon light is in the tacking field the tracking accuracy is about $\pm 5$ $\mu$ rad (RMS).

**Key factors for successful running of the experiments**

In order to maintain smooth running of the experiment several more essential factors should be addressed seriously. The first is the axiality for all the optical elements. We have first achieved axiality in the lab for polarization encoded module of 850 nm optical source, CCD for fine tracking, CCD for coarse tracking, and pulsed synchronization light, with the help of auxiliary mirror and optical source. Besides, one has to ensure stability before field experiments. In the receiver, axiality for optics is kept in the same way. The second is transmitter's divergence angles of 850 nm light and 532 nm light. In the experiment we have managed to make the divergence angle of 850 nm light as small as possible (restricted by diffraction-limited), and to keep the divergence angle of 532 nm light locating in a suitable level (~ 1 mrad) ensuring simultaneous operation for synchronization and ATP. Thirdly, through a three-dimension platform between optics and transmitter-telescope, not only can we instantly adjust the transmitter angle to a rough direction to maintain the beacon light of ATP within the effective field of view in the hot-air experiment, but also combining the camera set-up on the platform we can compensate in real-time the polarization-basis mismatch to restrict the QBER below acceptable threshold during the moving and floating experiments. Furthermore, polarization visibility is a very demanding parameter when we design the system. We have developed and used the following methods to improve our system's polarization visibility: All of the reflection mirrors are coated metal films for high polarization maintaining (above 1000:1); employing two mirrors for phase complement in polarization deflection; other optical elements (PBS etc.) are custom-made with high extinction rate. Considering all above features when setting up the optical system, we manage to achieve the polarization visibility of the transmitter system and receiver system to about 200:1, when transmitter and receiver are connected together in experiment, total polarization visibility is as high as 100:1.

## Acknowledgements

We are grateful to the staff of the Qinghai Lake National Natural Reserve Utilization Administration Bureau, especially Y.-B. He and Z. Xing, for their support during the experiment. This work was supported by Chinese Academy of Sciences, National Natural Science Foundation of China, and the National Fundamental Research Program (under grant no. 2011CB921300).

**Author Contributions** All authors contributed extensively to the work presented in this paper.

Correspondence and requests for materials should be addressed to J.-W.P. (pan@ustc.edu.cn) or C.-Z.P. (pcz@ustc.edu.cn).